\documentclass[onecolumn,showpacs,preprintnumbers,amsmath,amssymb]{revtex4}

\usepackage{graphicx}
\usepackage{dcolumn}
\usepackage{bm}

\begin{document}

\preprint{APS/123-QED}

\title{Spontaneous Symmetry Breaking Breaks Time-Reversal Symmetry}

\author{Jose A. Magpantay}
\email{jose.magpantay11@gmail.com}
\affiliation{Quezon City, Philippines}%

\date{\today}

\begin{abstract}
The ideas related to the arrow of time are discussed briefly. I then focus on the prevalent physical mechanism in the evolution of the universe and developments in particle physics, spontaneous symmetry breaking, and show that it explicitly breaks time-reversal symmetry. I do this in a point mechanics 'gauge theory', with 'gauge symmetry' under O(2). In this 'gauge model', I show that time reversal symmetry is broken in both classical and quantum physics because of the necessary use of the time step function when symmetries are spontaneously broken at a specific time in the system's evolution. I then argue that since spontaneous symmetry breaking is known to have happened once (electroweak breaking at $ t = 10^{-12} sec $  and maybe at least two more times during the earlier period in the universe's early evolution, then there should not be any more question about the arrow of time. Furthermore, the prevalence of time-irreversible processes at cosmic, galactic, stellar, planetary, biological and even societal levels, reflect the consistency with the microscopic irreversibility brought about by spontaneous symmetry breaking. 
\end{abstract}

\pacs{Valid PACS appear here}
\maketitle

\section{\label{sec:level1}Introduction}
The problem of the arrow of time stems from the observation that the physical laws like Newtonian dynamics and Maxwell's electromagnetism are time-reversal invariant while macroscopic systems are time irreversible. How can irreversibility of these macroscopic systems follow from microscopic laws that are reversible? Boltzmann showed that from microscopic laws that are time reversible, the (Boltzmann) entropy of a closed macroscopic system increases till it reaches equilibrium where it attains maximum value see any book on statistical mechanics, for example, \cite{Reichl}. This was used by Eddington \cite{Eddington} to argue that the arrow of time is essentially determined by entropy. 

But since entropy is a derived global, statistical quantity, it should not shield systems from pure dynamical results like that of Poincare Recurrence Theorem, which states that the system can eventually go back to its initial state, albeit after a really, really long time (much longer than the age of the universe), in the process going through a decrease in entropy along the way as pointed out by Zermelo. Loschmidt's criticism, on the other hand, strikes at the core of the problem, that the time-reversal symmetry of the basic equations does not preclude the system from reversing back to lower entropy state. These two issues have been discussed for a long time and it seems the consensus is in favor of Boltzmann \cite{Lebowitz}. However, as argued below, there is another way to resolve these issues, and it is to decouple entropy from the arrow of time, which will make the two criticisms of Boltzmann's ideas stand on their own dynamical merits.

The best description of the link between increase in entropy and forward flow of time was given by Muller (in a book for the layman) \cite{Muller}, it is at best a correlation. There is no known reason why entropy should cause the flow of time in a  particular direction. It is a fact that there are many open systems that have lowering of entropy for a certain period yet it is not claimed that the 'local' time of such a system goes backwards. If the 'local' time of an entropy reducing open system flows backwards while the rest of the environment have time flowing forward, it will create gradients in many parts of space-time especially in the vicinity of living systems. The gradients suggest curvature, does it mean then that space-time is no longer flat in the vicinity of living systems? It is to be noted that there are suggestions in the literature linking entropy and gravity \cite{Verlinde}, specifically that gravity is an entropic force. The case being considered here is biologically local suggesting that if indeed local decrease in entropy means curvature in space-time, how does it follow then from the idea of gravity as an entropic force? Or does it mean that the right hand side of Einstein's GR equations will have an extra term, aside from the energy momentum tensor, an entropy term? It seems that linking entropy to flow of time leads to questions that may need revision of established ideas on space-time. Maybe, it is better to unlink entropy and time  or at best as Muller suggests, it is merely a correlation.    
  
Philosophers proposed a solution to the time flow problem by proposing time has its own direction see for example \cite{Callender}. Time flowing forward only is interesting because the other physical problems like increasing/decreasing entropy, recurrence, collapse of the wave function, etc.,  must be settled within ideas in dynamics and not relate to the nature of time. This should have been the end of the issue but a time-ordering field, like a time potential, was introduced to account for the forward flow of time just like a gravitational potential accounts for up-down direction. This solution will open more problems, like a separate dynamics for time or explaining how the time potential couple to time asymmetric processes. Maybe, since the proposed solution is philosophical, it is sufficient to say time's nature is to flow forward - no need to explain but explore the philosophical consistency of time flowing forward and the inconsistency if time flows backwards. 

If we consider pure gravity, the diffeomorphism invariance of general relativity will make the Hamiltonian vanish, see for example \cite{Hanson} and by quantum mechanics, the wave function should be time-independent, time becomes irrelevant. We will not consider this any further because physics is not pure gravity so the Hamiltonian is never zero, time evolution is present as any astrophysical and cosmological observations show. 

A possible solution to the forward flow of time is microscopic irreversibility as explored in \cite{Magpantay}. A main source of this irreversibility is the physics of the early universe, in particular the spontaneous symmetry breaking that supposedly occurred when gravity decoupled from the other three forces at $ t = 10^{-43}  $ sec, when the other three forces separated into the strong and electroweak forces at $ t = 10^{-35}  $ sec, and finally when the electroweak forces separated into the weak and electromagnetic forces at $ t = 10^{-12} $ sec. Of the three phase transitions, the only one borne out by experiments at this time is the last. But this last phase transition introduced the overlooked irreversible aspect of Newtonian dynamics and Maxwell's electromagnetism. At $ t = 10^{-12} $ sec, the Higgs gave masses to the quarks and leptons, which slowed down the quarks enabling the gluons to capture them to form protons and neutrons with certain masses at $ t = 10^{-6} $ second . So the mass that appears in Newtonian dynamics came from the irreversible process that happened at that time. The same thing with the electric charge of the electron and proton, it came from the weak mixing angle taking on a definite value. Add to this the photon remaining massless while the weak bosons becoming massive when the phase transition happened and we see that the electromagnetic force of Maxwell has built in time-irreversible components even though the equations themselves are time reversal invariant. 

The shortcoming of the previous paper is that it was only argued that the spontaneous symmetry breaking led to the breakdown of time reversal symmetry. The argument being that there is no known example of a gauge symmetry being restored after symmetry breaking or a physical system naturally undoing its choice of direction in the internal symmetry space by reversing the flow of time. The same can be argued with inflation, the source of the bang in the Big Bang \cite{Guth}, because the early universe cannot roll back to the false vacuum, inflation cannot be uninflated. Still some believe that when the ultimate unification of all forces is achieved, it will be time-reversal invariant. Maybe, but why should it give spontaneous symmetry breaking in the very early universe where time-reversal symmetry is broken? Why can't the ultimate unification has built into it time-irreversibility? This cannot be settled until the ultimate unification is achieved. However, there is still the significant problem of showing that spontaneous symmetry breaking really breaks time-reversal symmetry. This is the problem addressed in this paper. 

The outline of the paper is then as follows. In Section II, I discuss an O(2) point mechanics 'gauge theory', which means the gauge parameter is time-dependent. I arranged the potential function so as to have the familiar shape, the wine bottle bottom, for spontaneous symmetry breaking. In Section III, I summarize the known facts about time-reversal invariance at the classical and quantum levels and show that spontaneous symmetry breaking breaks the time-reversal symmetry. This provides the proof of what was assumed in reference \cite{Magpantay}. I end in Section IV with a discussion of what was achieved in this paper and what it means for the realistic gauge theories with spontaneous symmetry breaking.    
 
\section{\label{sec:level2}Point Mechanics 'Gauge Theory'}
Consider a point particle in 2D, coordinates given by (x,y), coupled to a 2X2 'gauge' matrix $ A $ as described by the following Lagrangian
\begin{subequations}\label{1}
\begin{gather}
L = \frac{1}{2}\left( \dfrac{dA_{tr}}{dt} \right)^{2} + \frac{1}{2}\textbf{r}\overleftarrow{\left( \dfrac{d}{dt} + A^{T}(t) \right)}                                                              \cdot \overrightarrow{\left( \dfrac{d}{dt} + A(t) \right)} \textbf{r} - V(\textbf{r}),\label{first}\\
V(\textbf{r}) = \frac{k}{2}(x^{2} + y^{2}) + \frac{\lambda}{4} (x^{2} + y^{2})^{2},\label{second}\\
\textbf{r}(t) = \begin{pmatrix}
						x(t) \\
						y(t) 
						\end{pmatrix},\label{third}\\
A(t) = \begin{pmatrix}
			A_{11}(t) &  A_{12}(t) \\
			A_{21}(t) & A_{22}(t) 
			\end{pmatrix},
\end{gather}
\end{subequations}
where $ A_{tr} $ is the trace of the 2X2 matrix A, i.e., $ A_{tr} = A_{11} + A_{22} $, and $ A^{T} $ is the transpose of the matrix A. The point particle has unit mass, potentials that are harmonic with spring constant k and quartic with strength $ \lambda $.  
This Lagrangian is invariant under a time-dependent O(2) transformation given by
\begin{subequations}\label{2}
\begin{gather}
\textbf{r'}(t) = \Omega(t) \textbf{r}(t),\label{first}\\
A'(t) = \Omega(t) A(t) \Omega^{-1}(t) - \left( \dfrac{d\Omega}{dt} \right) \Omega^{-1},\label{second}\\
\Omega(t) = \begin{pmatrix}
							\cos \theta & \sin \theta \\
							-\sin \theta & \cos \theta 
							\end{pmatrix},
\end{gather}
\end{subequations}
where $ \theta = \theta (t) $. 
In component form, the O(2) gauge transformations given by equation (2) are
\begin{subequations}\label{3}
\begin{gather}
x' = x \cos \theta  + y \sin \theta ,\label{first}\\
y' = -x \sin \theta + y \cos \theta ,\label{second}\\
A'_{11} = A_{11}\cos ^{2}\theta + (A_{21} + A_{12}) \sin \theta \cos \theta + A_{22} \sin ^{2} \theta, \label{third}\\
A'_{22} = A_{11}\sin ^{2}\theta -(A_{21} + A_{12}) \sin \theta \cos \theta + A_{22} \cos ^{2} \theta,\label{fourth}\\
A'_{12} = (-A_{11} + A_{22})\sin \theta \cos \theta +A_{12} \cos ^{2}\theta - A_{21} \sin ^{2}\theta - \dot{\theta},\label{fifth}\\
A'_{21} = (-A_{11} + A_{22})\sin \theta \cos \theta - A_{12} \sin ^{2}\theta + A_{21} \cos ^{2}\theta + \dot{\theta}.
\end{gather}
\end{subequations}
Equations (3, c,d,e,f) show why the kinetic term of the gauge field is only expressed  in terms of $ \dot{A_{tr}} $, it is the only gauge-invariant term that can be formed with the matrix $ A(t) $. In general, any function of $ A_{tr} $ could have been added to the Lagrangian as this would have been gauge-invariant. But this would unnecessarily complicate the physics. 

For spontaneous symmetry breaking, it is the potential of $ \textbf{r} $, i.e., $ V =  \frac{k}{2}(x^{2} + y^{2}) + \frac{\lambda}{4}(x^{2} + y^{2})^{2} $ that is relevant. Equations (3,a,b) show that $ \vert\textbf{r}\vert = \sqrt{x^{2}+y^{2}} $ is gauge-invariant. Choosing $ k < 0 $, a repulsive harmonic potential, will give a wine bottle bottom shape for the potential  and the minimum is achieved when $ \vert\textbf{r}\vert = v = \sqrt{-\frac{k}{\lambda}} $, with the minimum potential value of $ -\dfrac{k^{2}}{2\lambda} $. The vacuum is then given by
\begin{equation}\label{4} 
\textbf{r}_{vac} = \begin{pmatrix}
									  v \cos \alpha (t) \\
									  v \sin \alpha (t)
									  \end{pmatrix},
\end{equation}
which shows the vacuum's continuous degeneracy. This suggests the following parametrization for $ \textbf{r} $
\begin{equation}\label{5}
\textbf{r} = \begin{pmatrix}
						\cos \alpha(t) & -\sin \alpha(t) \\
						\sin \alpha(t)  &  \cos \alpha(t)  
						\end{pmatrix}
						\cdot
						\begin{pmatrix}
						\eta + v \\
						0
						\end{pmatrix}
\end{equation}
Equation (5) transforms from degrees of freedom $ (x(t),y(t)) $ to $ (\eta(t), \alpha(t)) $. Under 'gauge' transformation defined in equation (2), $ \eta(t) $ is 'gauge'-invariant while the angular coordinate $ \alpha(t) $ goes into $ \alpha(t) - \theta(t) $.

Substituting equation (5) in equation (1), we find the following expression for the Lagrangian
\begin{equation}\label{6}
\begin{split}
L& = \frac{1}{2}\left(\dfrac{dA_{tr}}{dt}\right)^{2} + \frac{1}{2}\left(\dfrac{d\eta}{dt}\right)^{2} + \dfrac{d\eta}{dt}(\eta + v) \left[ \frac{1}{2}A_{tr} + \frac{1}{2}A_{d} \left( \cos ^{2}\alpha - \sin ^{2}\alpha \right) + \left( A_{12} + A_{21} \right)  \sin \alpha \cos \alpha \right]\\ 
& \quad + \dot{\alpha} (\eta + v)^{2} \left[ -A_{d} \sin \alpha \cos \alpha - A_{12} \sin ^{2}\alpha + A_{21} \cos ^{2}\alpha \right] + \frac{1}{2} (\dot{\alpha})^{2} (\eta + v)^{2} + \frac{1}{2}(\eta+v)^{2} \Bigl[\frac{1}{4} (A_{tr}-A_{d})^{2} + A_{tr}A_{d} \cos ^{2}\alpha + \\
& \quad + A_{12}^{2} \sin ^{2}\alpha + A_{21}^{2} \cos ^{2}\alpha + A_{tr} (A_{12}+A_{21} )\sin \alpha \cos \alpha + A_{d} (A_{12}-A_{21}) \sin \alpha \cos \alpha\Bigr] - \left[ \frac{k}{2} (\eta + v)^{2} + \frac{\lambda}{4} (\eta + v)^{4} \right] . \end{split}
\end{equation}
In the above equation, we identify $ A_{d} = A_{11} - A_{22} $. Note, the physical degrees of freedom, the gauge-invariant quantities, are $ \eta $ and $ A_{tr} $. 
This Lagrangian has three primary constraints, the conjugate momenta to $ A_{d} $, $ A_{12} $ and $ A_{21} $, which are $ \Pi_{d} $, $ \Pi_{12} $ and $ \Pi_{21} $ are all zeros. Carrying out Dirac's consistency formalism for this Lagrangian is too involved and will not elucidate the claim that spontaneous symmetry breaking breaks time-reversal symmetry. So, the thing to do is to simplify equation (6) but still have a remnant of gauge symmetry. 

This will be done in such a way that the gauge symmetry is simple and clear. This is done by taking
\begin{subequations}\label{7}
\begin{gather}
A_{11} = A_{22}, \label{first} \\
A_{12} = -A_{21},
\end{gather}
\end{subequations}
there are only two 'gauge fields', and the matrix is anti-symmetric. Note that under gauge transformation defined by equations (3; c, d, e, f), the structure defined by equation (7) is maintained. Using equation (7) in equation (6), the Lagrangian simplifies into
\begin{equation}\label{8}
\begin{split}
L_{s} & = \frac{1}{2} \left(\dfrac{dA_{tr}}{dt}\right)^{2} + \frac{1}{2} \left( \dfrac{d\eta}{dt} \right)^{2} + \frac{1}{2}\dfrac{d\eta}{dt}(\eta + v)A_{tr} + \frac{1}{8}(\eta + v)^{2} A_{tr}^{2} \\
& \quad - \left[ \frac{k}{2} (\eta + v)^{2} + \frac{\lambda}{4} (\eta + v)^{4} \right] + \frac{1}{2} (\eta + v)^{2} \left( \dot{\alpha} - A_{12} \right)^{2},
\end{split}
\end{equation}
where the subscript s in L means symmetric because there are still (a simplified) 'gauge' freedom, which are $ A_{12} \rightarrow A_{12} - \dot{\theta} $ (read from equations (3; c,d,e,f) ) and $ \alpha \rightarrow \alpha - \theta $ as discussed after equation (5), that leaves the Lagrangian $ L_{s} $ invariant. 

The classical equations of motion from $ L_{s} $ are given by
\begin{subequations}\label{9}
\begin{align}
\dfrac{d^{2}\eta }{dt^{2}} +\frac{1}{2} (\eta + v) \dfrac{dA_{tr}}{dt} - \frac{1}{4} (\eta + v) A_{tr}^{2} + \left[ k (\eta + v) +\lambda (\eta + v)^{3} \right]  - (\eta + v) \left( \dfrac{d\alpha }{dt} - A_{12} \right)^{2} &= 0, \label{first} \\
\dfrac{d^{2}A_{tr}}{dt^{2}} - \frac{1}{4} (\eta + v)^{2} A_{tr} - \frac{1}{2} \dfrac{d\eta }dt (\eta + v) &= 0, \label{second} \\
\dfrac{d\alpha}{dt}  - A_{12} &= 0, \label{third} \\ 
(\eta + v)^{2} \dfrac{d^{2}\alpha }{dt^{2}} + 2 \dfrac{d\alpha }{dt} (\eta + v) \dfrac{d\eta }{dt} - 2 A_{12} (\eta + v) \dfrac{d\eta }{dt} - (\eta + v)^{2} \dfrac{dA_{12}}{dt} &= 0.
\end{align}
\end{subequations}
Equation (9c) gives $ \dot{\alpha} = A_{12} $ and upon substituting in equation (9 d) yields a consistency resulting in $ A_{12}(t) $ undetermined, it is completely arbitrary. Equation (9a) becomes 
\begin{equation}\label{10}
\dfrac{d^{2}\eta}{dt^{2}} +\frac{1}{2} (\eta + v) \dfrac{dA_{tr}}{dt} - \frac{1}{4} (\eta + v) A_{tr}^{2} + \left[ k (\eta + v) + \lambda (\eta + v)^{3} \right] = 0,
\end{equation}
which along with equation (9 b) are the equations of motion in the symmetry broken phase, i. e., when the vacuum direction had been fixed at $ \alpha = 0 $. 

Thus, in the symmetric phase, the classical dynamics are comprised of the time evolution of the gauge-invariant variables $ (\eta(t), A_{tr}(t) $ as given by equations (10) and (9 b) plus the dynamics of the gauge-dependent quantities $ (\alpha(t), A_{12}(t) $, where $ A_{12}(t) $ is completely arbitrary and $ \alpha(t) $ determined by equation (9c). These, in turn will give $ (x(t), y(t)) $ as given by equation (5) while the gauge field $ A(t) $ is given by equation (1 d) with $ A_{11}(t) = A_{22}(t) = \frac{1}{2} A_{tr}(t) $ as given by equations (10) and (9 b) while $ A_{12} (t) = -A_{21}(t) $, which is completely arbitrary.  

Now, carry out the canonical quantization of the Lagrangian $ L_{s} $. It has a primary constraint given by
\begin{equation}\label{11}
\Pi_{12} = 0, 
\end{equation}
while the non-vanishing momenta, which will solve for the velocities are
\begin{subequations}\label{12}
\begin{gather}
\Pi_{tr} = \dot{A_{tr}}, \label{first} \\
\Pi_{\eta} = \dot{\eta} + \frac{1}{2} (\eta + v) A_{tr}, \label{second} \\
\Pi_{\alpha} = (\eta + v)^{2} \dot{\alpha} - (\eta + v)^{2} A_{12}.
\end{gather}
\end{subequations}
The Hamiltonian is given by 
\begin{equation}\label{13}
H_{s} = \frac{1}{2} \Pi_{tr}^{2} + \frac{1}{2} \Pi_{\eta}^{2} - \frac{1}{2} \Pi_{\eta} (\eta + v) A_{tr} + \left[ \frac{k}{2} (\eta + v)^{2} + \frac{1}{4} (\eta + v)^{4} \right] + \frac{1}{2} \dfrac{\Pi_{\alpha}^{2}}{(\eta + v)^{2}} + \Pi_{\alpha} A_{12}.
\end{equation}

Consistency of the primary constraint yields a secondary constraint
\begin{equation}\label{14}
\Pi_{\alpha} = 0,
\end{equation}
and the two constraints are first-class since 
\begin{equation}\label{15}
\left[ \Pi_{12}, \Pi_{\alpha} \right]_{PB} = 0.
\end{equation}
These first-class constraints $ \Pi_{12} $ and $ \Pi_{\alpha} $ are the generators of the gauge transformations $ A_{12} \rightarrow A_{12} - \dot{\theta} $ and $ \alpha \rightarrow \alpha - \theta $. As first-class constraints, they can then be taken as weakly equal to zero, which means the wave function $ \Psi(\eta, A_{tr}, \alpha, A_{12},t) $ of the Hamiltonian $ H_{s} $ is solved first, then the physical states are later determined by those that are annihilated by the first-class constraints, i.e., 
\begin{subequations}\label{16}
\begin{gather}
\hat{\Pi}_{12} \Psi = 0, \label{first} \\
 \hat{\Pi}_{\alpha} \Psi = 0.
 \end{gather}
 \end{subequations}
Thus, in the quantum theory of the symmetric phase, the state vectors are quite involved, which are solved from the Schroedinger equation using $ H_{s} $, then the physical states are projected out by equations (16; a,b). 

Suppose that at $ t = t_{0} $ the gauge symmetry is spontaneously broken, that is the vacuum direction is chosen, to say $ \alpha = 0 $ in equation (5) before substituting in equation (1) resulting in 
\begin{equation}\label{17}
L'_{sb}  = \frac{1}{2} \left(\dfrac{dA_{tr}}{dt}\right)^{2} + \frac{1}{2} \left( \dfrac{d\eta}{dt} \right)^{2} + \frac{1}{2}\dfrac{d\eta}{dt}(\eta + v)A_{tr} + \frac{1}{2}(\eta + v)^{2} \left[ \frac{1}{4}A_{tr}^{2} + A_{12}^{2} \right]  - \left[ \frac{k}{2} (\eta + v)^{2} + \frac{\lambda}{4} (\eta + v)^{4} \right]. 
\end{equation}
The gauge dependent quantity$ A_{12} $ is still present but easily removed in a few steps. The classical equation of motion of $ A_{12} $ simply gives $ A_{12} = 0 $ resulting in the equations of motion given by equations (9b) and (10) for the gauge-invariant quantities $ (\eta(t), A_{tr}(t)) $, which follows from the Lagrangian $ L_{sb} $ given by
\begin{equation}\label{18}
L_{sb} = \frac{1}{2} \left(\dfrac{dA_{tr}}{dt}\right)^{2} + \frac{1}{2} \left( \dfrac{d\eta}{dt} \right)^{2} + \frac{1}{2}\dfrac{d\eta}{dt}(\eta + v)A_{tr} + \frac{1}{8}(\eta + v)^{2} A_{tr}^{2}  - \left[ \frac{k}{2} (\eta + v)^{2} + \frac{\lambda}{4} (\eta + v)^{4} \right]. 
\end{equation} 

The classical dynamics is now given purely in terms of the gauge-invariant quantities, which although not easily solved, will no longer be complicated by the (infinite) multitudes of gauge-dependent terms $ (\alpha(t), A_{12}(t)) $. 

Doing a canonical quantization of equation (18) results in the Hamiltonian 
\begin{equation}\label{19}
H_{sb} = \frac{1}{2} \Pi_{tr}^{2} + \frac{1}{2} \Pi_{\eta}^{2} - \frac{1}{2} \Pi_{\eta} (\eta + v) A_{tr } + \left[ \frac{k}{2} (\eta + v)^{2} + \frac{\lambda}{4} (\eta + v)^{4} \right].
\end{equation}
Note, this Hamiltonian can also be derived from the Lagrangian given by equation (17) resulting in 
\begin{equation}\label{20}
H'_{sb} = H_{sb} - \frac{1}{2} (\eta + v)^{2} A_{12}^{2},
\end{equation}
but with a primary constraint $ \Pi_{12} = 0 $. This gives a secondary constraint $ A_{12} = 0 $. The two constraints are second class, thus must be imposed strongly equal to zero, which can be done by changing the Poisson brackets to Dirac brackets with only the Poisson bracket $ \left[ A_{12}, \Pi_{12} \right] $ changed to zero while the others are unchanged. This simplifies $ H'_{sb} $ to $ H_{sb} $. 

The quantum states derived from $ H_{sb} $ are all physical and do not have any complications of extra conditions like those given by equation (16).

These completes the classical and quantum descriptions of the symmetric and spontaneously broken phases of the point particle O(2) symmetric gauge model. 

\section{\label{sec:level3}Breakdown of Time-Reversal Symmetry}
The time-reversal invariance of physical systems follow from the fact that the starting point, the degrees of freedom in the Lagrangian L have appropriate behaviour under $ t \rightarrow -t $, which makes L invariant. This will then make the classical equations of motion also time-reversal invariant.  

In quantum mechanics, the starting point is the Schroedinger equation. Invariance under time-reversal of this equation requires that under $ t \rightarrow -t $, the Hamiltonian is invariant and the wave function $ \Psi(t) \rightarrow \Psi^{*}(-t) $. 

Inspecting the Lagrangians given by equations (6), (8), (16) or (17), they are all invariant under time-reversal if the fields transform under $ t \rightarrow -t $ as
\begin{subequations}\label{21}
\begin{gather}
\eta(t) \rightarrow \eta(-t) = +\eta(t), \label{first} \\
A_{ij}(t) \rightarrow A_{ij}(-t) = -A_{ij}(t), \label{second} \\
\alpha(t) \rightarrow \alpha(-t) = +\alpha(t),
\end{gather}
\end{subequations}
for $ i, j = 1,2 $, which makes $ A_{tr}, A_{d} $ also odd under time-reversal, while $ \eta $ and $\alpha $ are even. All the classical equations of motion using $ L_{s} $ or $ L_{sb} $ are also-invariant under time-reversal. Where is the breakdown of time-reversal then? It comes from the fact that the symmetry was spontaneously broken at $ t = t_{0} $, which means for $ t < t_{0} $ the Lagrangian of the system is given by $ L_{s} $ with its classical physics described in the discussions between equations (10) and (11) while for $ t > t_{0} $, the Lagrangian of the system is $ L_{sb} $ with its own classical physics purely in terms of the gauge-invariant quantities. 

The complete Lagrangian valid at any time t can be written down as
\begin{equation}\label{22}
L(t) = \Theta (t_{0} - t) L_{s} + \Theta (t - t_{0}) L_{sb},
\end{equation}
where $ \Theta (t) $ is the step function, with value equal to zero when $ t < 0 $, equal to 1 when $ t > 0 $. Note, the step function is a dimensionless quantity. Under time-reversal, all times will be reversed, including when the symmetry is spontaneously broken, i.e., $ t_{0} \rightarrow -t_{0} $,  $ t \rightarrow -t $. Both $ L_{s} $ and $ L_{sb} $ are time-reversal invariant as have been arranged in the previous section. But the step functions are not. For example, the first term $ \Theta (t_{0} - t) L_{s} \rightarrow \Theta (-t_{0} + t) L_{s} $. But $  \Theta (-t_{0} + t) $ is equal to one when $ t > t_{0} $, so for invariance under time-reversal this term must go with $ L_{sb} $ but it does not, it goes with $ L_{s} $. Similarly, $ \Theta (t - t_{0}) L_{sb} \rightarrow \Theta (-t + t_{0}) L_{sb} $. But $ \Theta (-t + t_{0}) $ is equal to one only when $ t < t_{0} $. But what goes with the $ \Theta (-t + t_{0}) $ is $ L_{sb} $, for invariance, it should have been $ L_{s} $. Thus, under time-reversal, $ L(t) $ given by equation (22) is not time-reversal invariant, in particular
\begin{equation}\label{23}
\Theta (t_{0} - t) L_{s} + \Theta (t - t_{0}) L_{sb} \rightarrow \Theta (t_{0} - t) L_{sb} + \Theta (t - t_{0}) L_{s},
\end{equation}
The same goes with the classical dynamics, it is not time-reversal invariant. In essence, what equation (23) says is that in the normal flow of time the system goes from the symmetric phase to the symmetry broken phase while the time reverse flow starts from symmetry broken phase to the symmetric phase. 

How about the quantum theory? First, similar to the Lagrangian, the Hamiltonian can be written as
\begin{equation}\label{24}
H(t) = \Theta (t_{0} - t) H_{s} + \Theta (t - t_{0}) H_{sb},
\end{equation}
where the explicit time-dependence of the Hamiltonian comes from the step function and both $ H_{s} $ and $ H_{sb} $ are real and time-reversal invariant. First off, the same analysis done on $ L(t) $ also holds for $ H(t) $, thus the Hamiltonian is not-time-reversal invariant and changes following equation (23). Secondly, since the Hamiltonian is not invariant under time reversal, it is not known how the wave function should transform for the Schroedinger equation to be invariant. For completeness, the solution to the Schroedinger equation with an explicitly time-dependent Hamiltonian is given by
\begin{subequations}\label{25}
\begin{gather}
\Psi (t) = U(t_{i}, t) \Psi (t_{i}), \label{first} \\
U(t_{i}, t) = \textbf{T}\left[ \exp {\left( -\frac{i}{\hbar} \int_{t_{i}}^{t} H(t') dt' \right) } \right],
\end{gather}
\end{subequations}
where $ t_{i} $ is the initial time, which we take to be less than $ t_{0} $ (otherwise, the system is already in the symmetry broken phase). The time-ordering operation $ \textbf{T} $ is given by
\begin{equation}\label{26}
U(t_{i}, t) =  1 + \frac{-i}{\hbar}  \int_{t_{i}}^{t} dt'_{1} H(t'_{1})  + (\frac{-i}{\hbar})^{2} \int_{t_{i}}^{t} dt'_{2} \int_{t_{i}}^{t'_{2}} dt'_{1} H(t'_{2}) H(t'_{1}) + (\frac{-i}{\hbar})^{3} \int_{t_{i}}^{t} dt'_{3} \int_{t_{i}}^{t'_{3}} dt'_{2} \int_{t_{i}}^{t'_{2}} dt'_{1} H(t'_{3}) H(t'_{2}) H(t'_{1}) + ....
\end{equation}
Substituting equation (24) and carefully tracking the integration times from $ t_{i} $ to $ t_{0} $ to $ t $ when appropriate, equation (26) simplifies into
\begin{equation}\label{27}
U(t_{i}, t) = \exp {\left( \frac{-i}{\hbar}[ H_{s}(t_{0} - t_{i}) + H_{sb} (t - t_{0}) ]\right) },
\end{equation}
where now the usual $ \frac{1}{n!} $ factors that go with the exponential expansion applies. The solution to the Schroedinger equation with the Hamiltonian given by equation (24) is given by equation (25a) with $ U(t_{i},t) $ given by equation (27). 

How about under time-reversal? In this case, since the Hamiltonian is not time-reversal invariant, the time-reversed wave function that satisfies the Schroedinger equation is not given by $ \Psi^{*})(-t) $. However, the time-reversed Hamiltonian is known and it is $  \Theta (t_{0} - t) H_{sb} + \Theta (t - t_{0}) H_{s} $. In the time-reversed sequence, the system begins at time t (symmetric phase) then passes through $ t_{0} $ to time $ t_{i} $ (symmetry broken phase). Thus, t and $ t_{i} $ exchanged roles with the solution given by
\begin{subequations}\label{28}
\begin{gather}
\Psi (t_{i}) = U_{r}(t, t_{i}) \Psi (t), \label{first} \\
U_{r}(t, t_{i}) =  \exp {\left( \frac{-i}{\hbar}[ H_{s}(t_{0} - t) ] + H_{sb} (t_{0} - t_{i}) ]\right)},
\end{gather}
\end{subequations}. 
where $ U_{r}(t, t_{i}) $ is the evolution operator in the time-reversed sequence. There is no point in comparing $ \Psi (t) $ in the forward direction to $ \Psi (t) $ in the time-reversed sequence because they are in different phases of the symmetry.

\section{\label{sec:level4}Conclusion} 
 I have shown, using a point mechanics model with an O(2) gauge symmetry, that spontaneous symmetry breaking breaks time-reversal symmetry. This is a simple model and the spontaneous symmetry breaking can be specified solely by the time when it happens. This leads to the use of the step function in time in distinguishing the symmetric phase from the spontaneously broken phase, which is significant in showing the breakdown of time reversal symmetry. 
 
 In real gauge theory, the symmetric phase is characterized by massless vector bosons , maybe massless fermions and scalars with negative mass. The spontaneously broken phase is characterized by massive vector bosons of the broken gauge symmetries resulting from the Goldstone bosons being eaten by the original massless vector fields, massless vector bosons of the remaining gauge symmetry, the fermions acquiring a mass from the Yukawa term and a physical massive scalar field, the so called Higgs boson. The difference between the two phases in real gauge theory are more dramatic than in the toy model considered in this paper. The history of the universe involves specifying the time, the size, temperature, matter density and energy scale of each era \cite{Collins}. Einstein's equations, thermodynamics and particle physics relate these parameters. Thus, specifying the phase transition is more involved. But to the extent that the time of the symmetry breaking is a significant factor in the phase transition justifies the use of the step function in time to write down a Lagrangian density that will cover the two phases. This is further supported by the fact that at the electroweak breaking scale, space-time can be considered flat and special relativity applies. Since the sign of $ t - t' $ is Lorentz invariant, i.e., causality is absolute, the Lagrangian density function written as
\begin{equation}\label{29}
\textit{L} = \Theta (t_{0} - t) \textit{L}_{s} + \Theta (t - t_{0}) \textit{L}_{sb},
\end{equation}
is Lorentz invariant. This expression says for times before the spontaneous breaking at $ t_{0} $, the system is in the symmetric phase and for times after $ t_{0} $, the system is in the symmetry broken phase. If the Lagrangians  $ \textit{L}_{s} $ and $ \textit{L}_{sb} $ are time-reversal invariant, unless there are CP violating terms which may come from the strong force (theta angle of QCD vacuum) and or the weak force (non-vanishing phases of the Cabibbo-Kobayashi-Maskawa matrix for the quarks and its counterpart for neutrinos, the Pontecorvo-Maki-Nakagawa-Sakata matrix) , then the proof of the breakdown of time-reversal symmetry follows from the $ \Theta $ term. Furthermore, the CP violations are weak and thus are not as significant as the breakdown of the time-reversal symmetry coming from spontaneous breaking of the gauge symmetry.  

To conclude, spontaneous symmetry breaking breaks time-reversal symmetry. As already discussed, this is known to have occurred at $ t = 10^{-12} $ second after the Big Bang and may have occurred at least two more times during the Universe's much earlier history. If time-reversal symmetry breakdown happened this early in the Universe's history, then at the microscopic level there is no time-reversal symmetry at all and there is no paradox of macroscopic irreversibility from microscopic reversibility. And maybe since spontaneous symmetry breaking happened very early in the Universe's history, the ultimate unification of all forces may have built into it the breakdown of time-reversal symmetry.  

\begin{acknowledgments}
I am grateful to Felicia Magpantay for correcting my Latex file. I would like to thank Gravity for patiently keeping me company while I worked on this paper. 
\end{acknowledgments}


\end{document}